# Survey of Server Virtualization

Radhwan Y Ameen
Department of Comp. Engineering
College of Engineering, Mosul University
Mosul, Iraq
radeeameen@gmail.com

Asmaa Y. Hamo
Dept. of Software Engineering
College of Computer Sc. and Mathematics, Mosul University
Mosul, Iraq
asmahammo@yahoo.com

*Abstract*— Virtualization is a term that refers to the abstraction of computer resources. The purpose of virtual computing environment is to improve resource utilization by providing a unified integrated operating platform for users and applications based on aggregation of heterogeneous and autonomous resources. More recently, virtualization at all levels (system, storage, and network) became important again as a way to improve system security, reliability and availability, reduce costs, and provide greater flexibility. Virtualization has rapidly become a go-to technology for increasing efficiency in the data center. With virtualization technologies providing tremendous flexibility, even disparate architectures may be deployed on a single machine without interference This paper explains the basics of server virtualization and addresses pros and cons of virtualization .

**Keywords- virtualization ,server ,hypervisor ,Virtual Machine Manager, VMM , para virtualization , full virtualization, OS level server.**

## I. INTRODUCTION

Virtualization is a technique for hiding the physical characteristics of computing resources from the way in which other systems, applications, or end users interact with those resources. It introduces a software abstraction layer between the hardware and the operating system and applications running on top of it [9] [ 1 ].This abstraction layer is called virtual machine monitor (VMM) or hypervisor and basically hides the physical resources of the computing system from the operating system (OS). Since the hardware resources are directly controlled by the VMM and not by the OS, it is possible to run multiple (possibly different) OSs in parallel on the same hardware. As a result, the hardware platform is partitioned into one or more logical units called virtual machines (VMs). "Virtuality" differs from "reality" only in the formal world, while possessing a similar essence or effect. In the computer world, a *virtual environment* is perceived the same as that of a *real environment* by application programs and the rest of the world, though the underlying mechanisms are *formally* different.

Virtualization was first developed in 1960's by IBM Corporation, originally to partition large mainframe computer into several logical instances and to run on single physical mainframe hardware as the host. This feature was invented because maintaining the larger mainframe computers became cumbersome. The scientist realized that this capability of partitioning allows multiple processes and applications to run at the same time, thus increasing the efficiency of the environment and decreasing the maintenance overhead[15].

## II. VIRTUAL MACHINE

### A. Virtual Machine History

Virtual machines have been in the computing community since 1960s, systems engineers and programmers at **Massachusetts Institute of Technology** (MIT ) recognized the need for virtual machines. In her authoritative discourse Melinda Varian [15] introduces virtual machine technology, starting with the ccompatible Time-Sharing System (CTSS).

IBM engineers had worked with MIT programmers to develop a time-sharing system to allow project teams to use part of the mainframe computers. Varian goes on to describe the creation, development, and use of virtual machines on the IBM OS/360 Model 67 to the VM/370 and the OS/390 [15]. Varian's paper covers virtual machine history, emerging virtual machine designs, important milestones and meetings, and influential engineers in the virtual computing community.

In 1973, Srodowa and Bates [14] demonstrated how to create virtual machines on IBM OS/360s. They describe the use of IBM's Virtual Machine Monitor, a hypervisor, to build virtual machines and allocate memory, storage, and I/O effectively. Srodowa and Bates touch on virtual machine topics still debated today: performance degradation, capacity, CPU allocation, and storage security.

Goldberg concludes "the majority of today's computer systems do not and cannot support virtual machines. The few virtual machine systems currently operational, e.g., CP-67, utilize awkward and inadequate techniques because of unsuitable architectures" [16].

Goldberg proposes the "Hardware Virtualizer," in which a virtual machine would communicate directly with hardware instead of going through the host software. Nearly 30 years later, industry analysts are excited about the announcement of hardware architectures capable of supporting virtual machines efficiently. AMD and Intel have revealed specifications for Pacifica and Vanderpool chip technologies with special virtualization support features.

The 1980s and early 1990s brought distributing computing to data centers. Centralized computing and virtual machine interest was replaced by standalone servers with dedicated functions: email, database ,Web, applications.
After significant investments in distributed architectures, renewed focus on virtual machines as a complimentary solution for server consolidation projects and data center management initiatives has resurfaced [17].

Recent developments in virtual machines on the Windows x86 platform merit a new chapter in virtual machine history. Virtual machine software from Virtuozzo, Microsoft, Xen, and EMC (VMWare) has spurred creative virtual machine solutions. Grid computing,computing on demand, and utility computing technologies seek to maximize computing power in an efficient, manageable way.

The virtual machine was created on the mainframe. It has only recently been introduced on the mid-range, distributed, x86 platform. Technological advancements in hardware and software make virtual machines stable, affordable, and offer tremendous value, given the right implementation.

*B. Virtual Machine Concepts*

Goldberg R. P defined Virtual machines as :"A system...which...is a hardware-software duplicate of a real existing machine, in which a non-trivial subset of the virtual machine's instructions execute directly on the host machine..." [22,23].While Goldberg R, June defined Virtual machines as:
"A virtual machine is taken to be an *efficient, isolated duplicate* of the real machine. We explain these notions through the idea of a *virtual machine monitor"* (VMM).

See Figure 1.

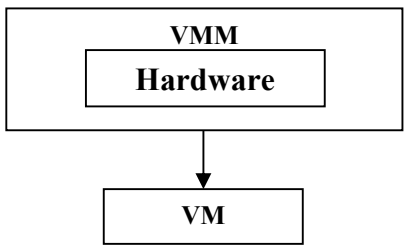

Fig. 1 The virtual machine monitor

As a piece of software a VMM has three essential characteristics.

First, the VMM provides an environment for programs which is essentially identical with the original machine;

second, programs run in this environment show at worst only minor decreases in speed; and last, the VMM is in complete control of system resources".[ 20] and Kreuter, D defined it as:

A virtual machine (VM) is an abstraction layer or environment between hardware components and the end- user. Virtual machines run operating systems and are sometimes referred to as virtual servers. A host operating system can run many virtual machines and shares system hardware components such as CPUs, controllers, disk, memory, and I/O among virtual servers" [18].

*C. Virtual Machine Types*

Virtual machines are implemented in various forms. Mainframe, open source, para virtualization, and custom approaches to virtual machines have been designed over the years. Complexity in chip technology and approaches to solving the x86 limitations of virtualization have led to three different variants of virtual machines:

**1.** software virtual machines (see Figure 2), which manage interactions between the host operating system and guest operating system (e.g., Microsoft Virtual Server 2005);

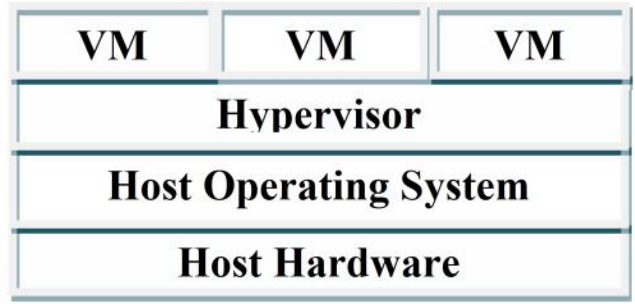

Fig. 2 Software virtual machines

**2.** hardware virtual machines (see Figure 3), in which virtualization technology sits directly on host hardware (bare metal) using hypervisors, modified code, or APIs to facilitate faster transactions with hardware devices (e.g., VMWare ESX);

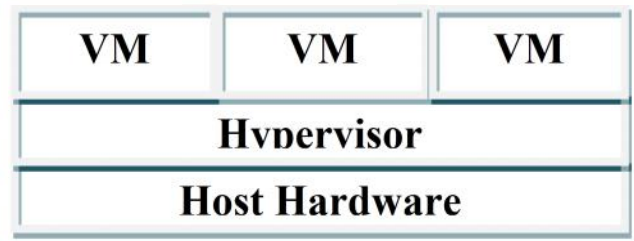

Fig. 3 Hardware virtual machines.

**3.** virtual OS/containers (see Figure 4), in which the host operating system is partitioned into containers or zones (e.g., Solaris Zones, BSD Jail).

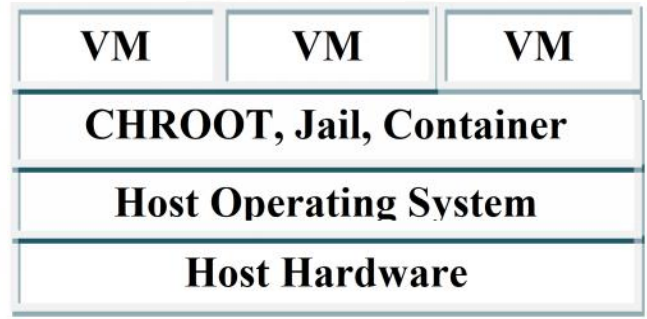

Figure 4: Virtual OS/containers virtual machines.

A simple UNIX implementation called *chroot* allows an alternate directory path for the root file system. This creates a "jail," or sandbox, for new applications or unknown applications. Isolated processes in chroot are best suited for testing and applications prototyping. They have direct access to physical devices, unlike emulators. Sun Microsystems' "Solaris Zones" technology is an implementation of chroot, similar to the FreeBSD jail design, with additional features. Zones allow multiple applications to run in isolated partitions on a single operating system [19].

Each zone has its own unique process table and management tools that allow each partition to be patched, rebooted, upgraded, and configured separately. Distinct oot privileges and file systems are assigned to each zone. [20].

### III. VIRTUALIZATION

#### A. Definitions of Virtualization

There are many definitions of term virtualization as shown below:

Rune Johan Andresen defined it as : "*Virtualization* is a framework of dividing the resources of a computer into multiple execution environments. More specific it is a layer of software that provides the illusion of a real machine to multiple instances of virtual machines." [4] [11]

While Susanta Nanda , Tzi-cker Chiueh defined it as : "Virtualization is a technology that combines or divides computing resources to present one or many operating environments using methodologies like hardware and software partitioning or aggregation, partial or complete machine simulation, emulation, time-sharing, and many others". . [7]

IBM defined it as: "*Virtualization* is the creation of substitutes for real resources, that is substitutes that have the same functions and external interfaces as their counterparts, but that differ in attributes, such as size, performance, and cost." [29]

Andi Mann defined it as : "Virtualization is, at its foundation, a technique for hiding the physical characteristics of computing resources from the way in which other systems, applications, or end users interact with those resources. This includes making a single physical resource (such as a server, an operating system, an application, or storage device) appear to function as multiple logical resources; or it can include making multiple physical resources (such as storage devices or servers) appear as a single logical resource." . [9]

G. Heiser defined it as:"virtualization allows a single computer to host multiple virtual boards (or virtual machines), each isolated from one another, with the possibility of running different operating systems. The main advantage is that, if a virtual board fails, the other ones are kept safe at a reasonable cost "[10].

William von H defined it as : "Virtualization is simply the logical separation of the request for some service from the physical resources that actually provide that service". . [8]

Chaudhary V.,Minsuk Cha.,Walters J.P.,Guercio S.,Gallo S, defined it as :"Virtualization is a common strategy for improving the utilization of existing computing resources, particularly within data centers."[3]

Amit Singh defined it as: "Virtualization is framework or methodology of dividing the resources of a computer into multiple execution environments, by applying one or more concepts or technologies such as hardware and software portioning, time-sharing, partial or complete machine simulation, emulation, quality of service, and many others."[2]

*Joshua S. White,Adam W. Pilbeam* defined it as: "Virtualization is a mechanism permitting a single physical computer to run sets of code independently and in isolation from other sets". [6]

Sahoo J., Mohapatra S., Lath R. defined it as: "Virtualization is a technology that introduces a software abstraction layer between the hardware and the operating system and applications running on top of it."[l]

TBD Networks defined it as: "Virtualization is a technology that enables running two or more operating systems simultaneously on a single computer."[5]'

And Lawrence C. Miller, CISSP defined it as: *"Virtualization is* technology emulates real or physical computing resources, such as desktop computers and servers, processors and memory, storage systems, networking, and individual applications."[25]

We define it as: "virtualization is a technology to divided or combined the resources of computer system between multiple operating systems or applications, to make illusion that each one access the real resources".

#### B. Benefits of virtualization

There can be innumerous reasons how virtualization can be useful in practical scenarios, a few of which are the following:

- Server Consolidation. [9] [23] [ 8] [25] [29] [40]
- Application consolidation. [ 8] [25] [40]
- Sandboxing[ 8,].
- Multiple execution environments[ 8] [40]
- Virtual hardware. [ 8]
- Multiple simultaneous OS[13]. [ 8, [40].
- Debugging. [ 8]
- Software Migration. [ 8]
- Appliances[23] [ 8].
- Testing[23] [40].
- Better Use of Existing Hardware[ 9].
- Reduction in New Hardware Costs[9] [8][23][28] [36]
- Reduction in IT Infrastructure Costs[ 9] [23] [36]
- Reduced downtime[9] [28] [36].
- Simplified System Administration[ 9].

- Increased Uptime and Faster Failure Recovery[ 9]
- Simplified Capacity Expansion[ 9]
- Simpler Support for Legacy Systems and applications [12] [ 9] [23] [40].
- Simplified System - Level Development[ 9] [40].
- Simplified System Installation and Deployment[ 9].
- Simplified System and Application Testing Business Continuity and Disaster Recovery[9] [ 9] [23] [25] [28] [29].
- Business Agility[9].
- Resource sharing .
- Isolation [13]. [12].
- Increase Flexibility. [9] [13]. [40]
- Increase Availability[23] [26] [36].
- Increase Scalability[23] [26] [36] .
- Increase Hardware utilization[12]. [26]
- Increase Security[12]. [26] [.
- Load Balancing[36]
- brings hardware independence[13] [26].

C. *Disadvantage of virtualization*
- SPOF Single Point of Failure Problem.
- Overhead causing decreased performance has been the biggest con with virtualization.
- The management interface This can be a problem as it encumbers consolidation of several platforms into the same environment.
- Increase in Networking Complexity and Debugging Time. [1][8][9].

D. *Types of virtualization*

There are so many different types of virtualization, Mobile, Data, Memory, Desktop, Storage, Server, Network, Application, Grid, and Clustering as shown in fig 5.

*1) Mobile Virtualization*

VMware defined it as : Mobile Virtualization(MVP) is a thin layer of software that is embedded on a mobile phone to decouple the applications and data from the underlying hardware. It is optimized to run efficiently on low power consuming and memory constrained mobile phones. The MVP currently supports a wide range of real-time and rich operating systems including Windows CE 5.0 and 6.0, Linux 2.6.x, Symbian 9.x, eCos, μITRON NORTi and μC/OS-II.

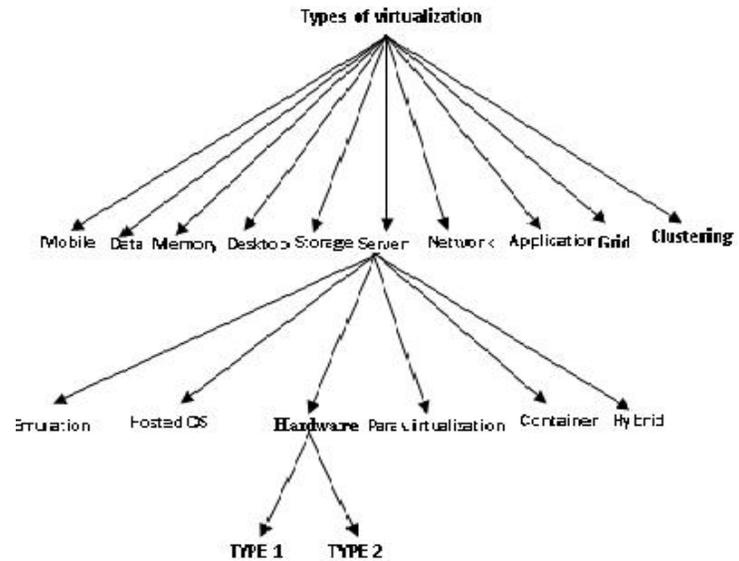

Fig. 5 Types of virtualization

*2) Data Virtualization*

Data virtualization from Andi Mann: "Data virtualization abstracts the source of individual data items including entire files, database contents, document metadata, messaging information, and more and provides a common data access layer for different data access methods such as SQL, XML, JDBC, File access, MQ, JMS, etc. This common data access layer interprets calls from any application using a single protocol, and translates the application request to the specific protocols required to store and retrieve data from any supported data storage method. This allows applications to access data with a single methodology, regardless of how or where the data is actually stored." [9]

*3) Memory Virtualization*

Carl A. Waldspurger ,Palo Alto define it as:

A guest operating system that executes within a virtual machine expects a zero-based physical address space, as provided by real hardware. ESX Server gives each VM this illusion, virtualizing physical memory by adding an extra level of address translation. Borrowing terminology from Disco [34], a machine address refers to actual hardware memory, while a physical address is a software abstraction used to provide the illusion of hard-ware memory to a virtual machine. We will often use "physical" in quotes to highlight this deviation from its usual meaning.[33]

*4) Desktop Virtualization*

William von H defined it as: The term "desktop virtualization" describes the ability to display a graphical desktop from one computer system on another computer system or smart display device.

Many window managers, particularly those based on the X Window System, also provide internal support for multiple, virtual desktops that the user can switch between and use to display the output of specific applications. The X Window System also supports desktop virtualization at the screen or display level, enabling window managers to use a display region that is larger than the physical size of your monitor.[8]

*5) Storage Virtualization*

Li Bignag, Shu Jiwu, Zheng Weimin defined it as : "Storage Virtualization is the emerging technology that creates logical abstractions of physical storage systems. Storage Virtualization has tremendous potential for simplifying storage administration and reducing costs for managing diverse storage assets."[21]

*6) Network Virtualization*

Naga Dinesh define it as: *"Network Virtualization* :would provide abstraction layer that can decouple the physical network equipment from the delivered business services over the network to produce a more responsive and well-organized communications"[22]

*7) Application Virtualization*

Naga Dinesh Defined it as*: "*This type of virtualization allows the user to run the application using local resources without installing the application in his system completely".[22]

While Joshua S. White, Adam W. Pilbeam define it as *:* "provides smaller single application virtual machines that allow for emulation of a specific environment on a client system. For example a Java Virtual Machine allows disparate operating systems such as Windows and Linux to run the same Java program as long as they have the Java VM installed. This form of virtualization is limited in that it only provides single program isolation from the host, but is useful when testing programs out without installing them". [6]

*8) Grid Computing*

Andi Mann defined it as: "Like a cluster, a grid provides a way to abstract multiple physical servers from the application they are running. The major difference is that the computing resources are normally spread out over a wide network, potentially across the Internet, and the physical servers that comprise a grid do not have to be identical. Unlike a cluster, where each server is locally connected, is likely to be identical, and can handle the same processing requirements, a grid is made up of heterogeneous systems, in diverse locations, each of which may specialize in a particular processing capability. Much greater coordination is needed to allocate the resources to appropriate workloads." [9]

*9) Clustering*

Andi Mann define it as: "A cluster is a form of virtualization that makes several locally-attached physical systems appear to the application and end users as a single processing resource. This differs significantly from other virtualization technologies, which normally do the opposite, i.e. making a single physical system appear as multiple independent operating environments. A typical use case for clustering is to group a number of identical physical servers to provide distributed processing power for high-volume applications, or as a "Web farm", which is a collection of Web servers that can all handle load for a Web-based application." [9]

*10) Server Virtualization( machine , cpu )*

The terms " server virtualization " , " machine virtualization "and "cpu virtualization" describe the ability to run an entire virtual machine, including its own operating system, on another operating system. The most common virtualization known in general is Server Virtualization.

*a) Server Virtualization definitions:*

Lawrence C. Miller defined it as: "*Server virtualization* creates "virtual environments" that allow multiple applications or server workloads to run on one computer, as if each has its own private computer".[25]

While Naga Dinesh define it as: "The technique of masking of server resources, which includes the identity and number of every existing servers, processors, and OS users is termed as server virtualization".[22]

HP define it as: "*server virtualization* refers to abstracting, or masking, a physical server resource to make it appear different logically to what it is physically. In addition, server virtualization includes the ability for an administrator to relocate and adjust the machine workload."[26]

VMware define it as: "virtualization enables one computer to perform the job of multiple computers, by sharing the resources of a single computer across multiple environments".[30]

Citrix system define it as: "the ability to decouple software from the hardware layer, allowing server workloads to be streamed onto any platform in any direction." [32]

And Darla Sligh define it as: *"Server virtualization* is a software-based tool enabling the division of computer resources and the sharing of multiple environments simultaneously." [31]

We define it as: "server virtualization is the ability to run many operating systems with isolation and independences on other operating system"

*b) Server Virtualization types*

Figure 6 show the traditional computer system without virtualization.

In x86 environments, there are several variations within software-layer abstraction of the server hardware, including these general categories:

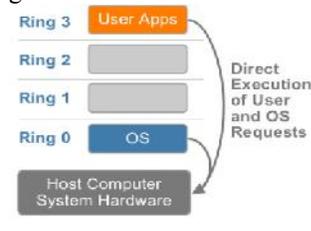

Fig. 6  x86 privileged level architecture without virtualization.

*i. Emulation*

Emulation is a virtualization method in which a complete hardware architecture may be created in software. This software is able to replicate the functionality of a designated hardware processor and associated hardware systems. This method provides tremendous flexibility in that the guest OS may not have to be modified to run on what would otherwise be an incompatible architecture. Emulation features tremendous drawbacks in performance penalties as each instruction on the guest system must be translated to be understood by the host system. This translation process is extremely slow compared to the native speed of the host, and therefore emulation is really only suitable in cases where speed is not critical, or when no other virtualization technique will serve the purpose. Examples of this approach are QEUM, Bochs , crusoe , and BIRD. [6] [7][8]

Advantages:
- tremendous flexibility in that the guest OS may not have to be modified.

Disadvantages:
- performance penalties as each instruction on the guest system must be translated to be understood by the host system.

*ii. Binary translation*

With binary translation technology as shown in figure 7, the guest OS is not aware it is operating on virtualized hardware. The hypervisor manages the access of each guest OS to the physical hardware resources by masking the hardware from the guest OS. It emulates portions of the system hardware and provides the guest OS with the illusion of a standard physical server with well-defined hardware devices. The hypervisor ensures that any instructions from the guest OS that affect system parameters—such as privileged instructions to the CPU—are handled in a way that does not affect the operation of other guest operating systems or cause OS kernel faults. The hypervisor traps the instruction and performs necessary translations that make the guest OS think it has complete control over the server hardware. The critical issue of dynamical binary translation is its low performance efficiency and design complexity due to the incapability of classical trap-and-emulate virtualization with previous generation of x86 architecture. Examples of this approach is VMWare ESX [26][31][8][39] [38]

Advantages:
- provides the guest OS with the illusion of a standard physical server with well-defined hardware devices.
- No need to modified guest OS.

Disadvantages:
- low performance efficiency
- design complexity due to the incapability of classical trap-and-emulate virtualization

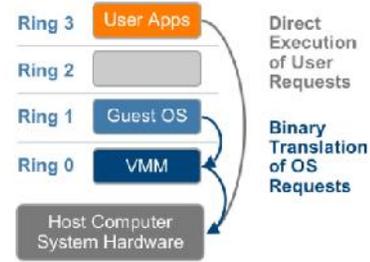

*Fig.* 7 The binary translation approach

*iii. Hosted OS, application-layer abstraction virtualization*

With Hosted OS, application-layer abstraction virtualization as shown in figure 8,another software-only approach uses a hypervisor layer that is hosted by an underlying OS. Because it resides as an application on top of the host OS, this type of abstraction inherits its hardware support and device compatibility from the host OS. This provides an advantage for customers who want to run an older, legacy OS on newer server hardware. However, the tradeoff for this hardware compatibility is the performance overhead required by the hypervisor layer. Typically, such hosted solutions are used in smaller, departmental environments rather than in large data center deployments because the hosted solutions often lack capabilities such as dynamic load balancing or clustering. Examples of this approach are Microsoft Virtual Server 2005 R2and VMware Server (formerly VMware GSX)[26][31] [35][8][ 37] [38].

Advantages:
- Virtualization product is installed onto the host desktop just as any other application
- The host desktop OS can continue to be used
- Uses the host OS's device drivers - the virtualization product supports whatever hardware the host does

Disadvantages:
- Slow performance. [36]

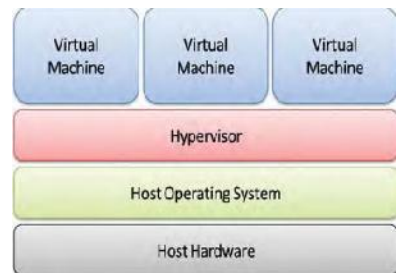

*Fig* 8 Hosted OS, application-layer abstraction virtualization

*iv. Hardware- assisted virtualization (full virtualization, bare-metal virtualization)*

With hardware-assisted virtualization (sometimes referred to as full virtualization) as shown in figure 9, the hypervisor is assisted by the processor hardware such as AMD-V or Intel VT-x processor virtualization technologies. In this scenario, when the guest OS makes a privileged instruction call, the processor (CPU) traps the instruction and returns it to the hypervisor to be emulated. Once the operation is serviced by means of the hypervisor, the modified instruction is returned back to the CPU for continued execution. Hardware assistance reduces the software overhead required by the hypervisor. Hardware assistance from AMD-V and Intel VT-x technologies extends the x86 instruction set with new instructions that affect the processor, memory, and local I/O address translations. The new instructions enable guest operating systems to run in the standard Ring-0 architectural layer, as they were designed to do, removing the need for ring compression. Examples of this approach are Microsoft Hyper-v, Citrix Xen , Parallels Workstation, Virtual Iron and VMWare ESX Server [26][31][1][7][8][35] [ 37] [38]

Advantages:
- Performance .
- Products are distributed as appliances or server OSes.

Disadvantages
- Vendor publishes a hardware compatibility list (HCL) that dictates what hardware can be used with their virtualization product. [36]

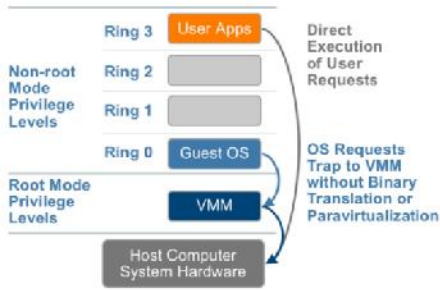

*Fig.* 9  The hardware assist approach

v. *Paravirtualization*

Paravirtualization as shown in figure 10, refers to a technique in which the guest OS includes modified (paravirtualized) I/O drivers for the hardware. Unlike a binary translation approach, the hypervisor does not need to trap and translate all privileged layer instructions between the guest OS and the actual server hardware. Instead, the modified guest OS makes calls directly to the virtualized I/O services and other privileged operations. Therefore, paravirtualization techniques have the potential to exhibit faster raw I/O performance than binary translation techniques. Some of the hypervisor implementations that use this method (Citrix XenServer, Red Hat Enterprise Linux 5, and SUSE Linux Enterprise) are unique in that they support paravirtualization when using a modified guest OS and hardware-assisted virtualization when the guest OS is not virtualization-aware. Device interaction in paravirtualized environment is very similar to the device interaction in full virtualized environment; the virtual devices in paravirtualized environment also rely on physical device drivers of the underlying host. Where paravirtualization differs is that it does not simulate hardware resources but instead offers a special Application Programming Interface (API) to hosted virtual machines. Examples of this approach are Xen, Denali and User-Mode Linux (UML) [36] [26][31][1][6] [3][7] [35] [ 37] [38].

Advantages:
- significant performance improvements over other virtualization solutions

Disadvantages:
- The VM OS must be modified.

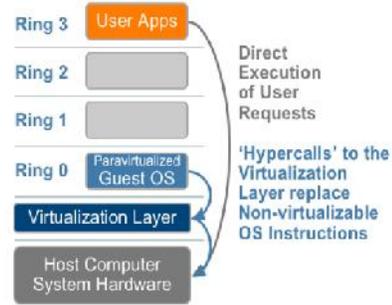

*Fig.* 10  The Paravirtualization approach

vi. *Hosted OS, kernel-layer abstraction (OS Containers virtualization, Single Kernel Image (SKI))*

Kernel-layer abstraction as shown in figure 11, refers to a technique in which the abstraction technology is built directly into the OS kernel rather than having a separate hypervisor layer. System - level virtualization is based on the change root (CHROOT) concept that is available on all modern UNIX - like systems.. The direct access to hardware could potentially provide greater performance than using a binary translation technology; however, because there is no separation between the hypervisor and the operating system, there is the possibility that resource conflicts may occur between multiple virtual machines. Virtual OS containers do not use hypervisors (or VMM), which is a software application that works to manage the logical separate of physical resource [40]. They use containers, or sandboxes, called chroot, to partition the host operating system into containers or zones (e.g., Solaris Zones, BSD Jail), so multiple applications can run in isolated partitions on a single operating system. [26][31] [ 35] [ 37] this concept implements virtualization by running more instances of the same OS in parallel. This means that not the hardware but the host OS is the one being virtualized[1] OS-layer virtualization tends to be more efficient and fails only by little to provide the same isolation [41]. Examples of this approach are FreeBSD ' s chroot jails, FreeVPS, Linux VServer, OpenVZ, Solaris Zones and Containers, and Virtuozzo. [8] [38].

Advantages:

- Performance
- Reduced disk space requirements, containers can use the same files

Disadvantages:
- The VM OS must be the same OS as the host OS. [36]

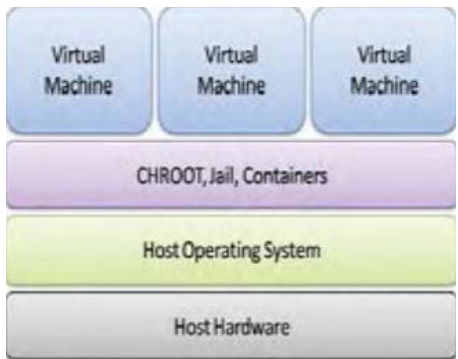

*Fig.*11  Containers virtualization

    vii. *Native virtualization, Hybrid virtualization, a hybrid*

virtualization as shown in figure 12, is the newest form. It is a combination of full virtualization and paravirtualization and uses input/output (I/O) acceleration techniques. This compromise allows for an increase in speed (and indeed with hardware acceleration it can be very fast), but potential performance degradation can exist in an environment where the instructions are relying more heavily on the emulated actions rather than the direct hardware access portions of the hypervisor It adds overhead and complexity. Examples of this approach are VMware and Microsoft Virtual PC [31] [6][35] [38]

Advantages:
- Performance

Disadvantages:
- Requires the underlying processor have virtualization extensions (examples: Intel-VT, AMD-V) to function.
- Older hardware that could otherwise be utilized by other virtualization architectures cannot be used. . [36]

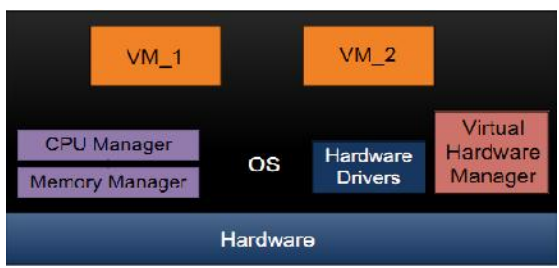

*Fig.* 12  **The Hybrid Virtualization**

IV.   TYPES OF HARDWARE VIRTUALIZATION

1) *Type 1(native or bare metal ).*
2) *Type2(hosted ).*

The kernel was known as the *supervisor* in mainframes; hence the term *hypervisor* was coined for the software operating above the supervisor.

Two types of hypervisors are defined for server virtualization:

Type 1 and Type 2 (see Figure 13,14). A Type 1 hypervisor, also known as a *native* or *bare metal* hypervisor, type 1 hypervisors run directly on the system hardware. The following figure shows one physical system with a type 1 hypervisor running directly on the system hardware, and three virtual systems using virtual resources provided by the hypervisor.

A Type 2 hypervisor, also known as a *hosted* hypervisor, it run on a host operating system that provides virtualization services, such as I/O device support and memory management. The following figure shows one physical system with a type 2 hypervisor running on a host operating system and three virtual systems using the virtual resources provided by the hypervisor.[25][29]

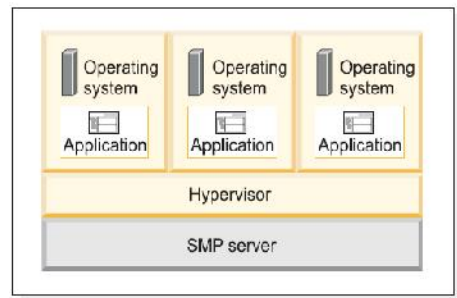

*Fig. 13* Type 1 hypervisors.

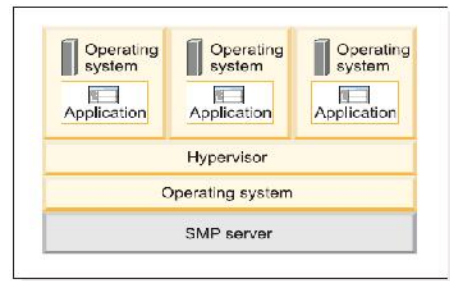

*Fig.14* Type 2 hypervisors

c) *Advantages of server virtualization*

many researchers note the following benefits for virtualizing servers within data centers

- enabling automated data center operations[24]
- improving the speed of service delivery[24]
- supporting application configuration and availability. [29]
- Consolidation to reduce hardware cost
- reducing the need for physical servers[24].
- reducing server operational maintenance[28].
- reducing the operating expense][26].
- reducing provisioning and the deploying of new services[28]
- reducing disaster recovery times[24] ][26]
- improving network and application security[27]

- reducing costs associated with the test and development of in-house applications[27] [29]

- enabling simple, responsive, utility-style „cloud computing‟ infrastructure[27]

- reducing various testing and migration issues][26].

- Reducing (TCO) Total Cost of Ownership[27] [29]

- Improving Flexibility, High Availability and Performance[27] [29]

*d) Pitfalls of server virtualization*

Researchers indicate that improper employment of server virtualization can result in the following pitfalls

- overloading the server utilization infrastructure, which can introduce application latency; [31]

- increasing IT operational costs because of additional time and resources required for extensive research efforts; [31]

- magnifying failures because a hardware failure could impact multiple virtual servers and the applications they host;[31]

- introducing virtual machine sprawl, which may substantially increase the overall number of server operating images that need to managed by system administrators; [31]

- enabling improper security processes because within the virtual server, the server administrator with access to the root ID can alter or disable security settings; thereby , exposing servers to security vulnerabilities;[31]

- exposing IT operations to network (traffic) uncertainties][31]

- requiring enhanced IT skill sets to manage more environments at once.[31]